\documentclass[preprint,showpacs,preprintnumbers,amsmath,amssymb]{revtex4-1}

\usepackage[pdftex]{color,graphicx}
\usepackage{dcolumn}% Align table columns on decimal point
\usepackage{bm}% bold math

\begin{document}

%\preprint{APS/123-QED}

\title{Application of the Space-Time Method to Stimulated Raman Adiabatic Passage on the Simple Harmonic Oscillator}

% Solve Time Dependent Schr{\"o}dinger Equation
\author{X.~Zhang and C.~Weatherford}
\affiliation{Dept.\ of Physics; Florida A\&M University;
Tallahassee, FL 32310}
\email{Charles.weatherford@famu.edu}
\homepage{}

\date{\today}% It is always \today, today,

\begin{abstract}
The space-time method is applied to a model system-the Simple Harmonic Oscillator in a laser field to simulate the Stimulated Raman Adiabatic Passage (STIRAP) process. The Space-Time method is a computational theory first introduced by Weatherford et. al. to solve Time-Dependent Systems with one boundary value and applied to electron spin system with invariant Hamiltonian [Journal of Molecular Structure {\bf 592} 47]. The implementation in the present work provides an efficient and general way to solve the Time-Dependent Schr{\"o}dinger Equation and can be applied to multi-state systems. The algorithm for simulating the Simple Harmonic Oscillator STIRAP can be applied to solve STIRAP problems for complex systems.
\end{abstract}

%\begin{abstract}
%\end{abstract}
%\pacs{}
%\pacs{ }
%\keywords{}
\maketitle
\section{Introduction}

In the past years, many numerical methods have been developed to solve the Time-dependent Schr{\"o}dinger Equations (TDSE) \cite{CLEFORESTIERJCOMPPHY94}. Weatherford et al introduced a Finite-Element Space-Time Algorithm and applied it to a system which is the procession of a single electron spin in constant magnetic field \cite{CAWEATHERFORDJMS592}. Later, Gebremedhin and Weatherford et al used the algorithm to evaluate the exponential of a matrix, they claimed that the algorithm can be treated as a black box since it is not matrix dependent \cite{DGEBREMEDHINArX0811}. 

The Finite-Element Space-Time Algorithm has not been used to solve a real time-dependent system ever since it was introduced by Weatherford \cite{CAWEATHERFORDJMS592}. In their work, an explicit description of the how to use a space-time basis set is given, but the Hamiltonian of the model system is time-independent. The problem solved in their work is a two-state problem. In the present work, the model system is a simple harmonic oscillator in laser field, thus the Hamiltonian is time-dependent and the system considered can have multiple states. Calculation results are shown for systems with two states, three states and more states. 

We also calculated population transfer of adiabatic passage for two-level systems and stimulated Raman adiabatic passage (STIRAP) for three-level systems. The model system used in this work is a one dimensional simple harmonic oscillator with an external laser field applied to it. Our goal is to verify and obtain some variable requirements for full population transfer with resonant exitation. Our computing process will provide a way for choosing the amplitudes and pulse length of pump pulse and stokes pulse in STIRAP process for other systems. We use time-basis projection method to solve the time-dependent Schrodinger equation. We present our results for different external laser fields. The results for Gaussian pulses laser field may provide a way for experiments.

This work is arranged in five sections. The Finite-Element Space-Time Algorithm with time-dependent Hamiltonian are introduced in section II. In section III, the algorithm is applied to the simple harmonics oscillator model system. In section IV, calculations are done to show that the algorithm works well for the model system. The effects of parameter value change are also presented. Section V is conclusions and analysis.

\section{Finite-Element Space-Time Algorithm with Time-Dependent Hamiltonian}

For a time-dependent quantum system, the Schr{\"o}dinger Equation (TDSE) is given by

\begin{equation}
\label{eq:ETDSET}
i\hbar\frac{\partial}{\partial t} \Psi \left(\vec{x},t \right) =\widehat{\cal H } \left( \vec{x} , t \right) \Psi \left( \vec{x} , t \right).
\end{equation}

The Hamiltonian ${\cal H}$ is,
\begin{equation}
\widehat{\cal H}\left( \vec{x} , t \right) = {\widehat{\cal H}}_0 \left( \vec{x} \right) + V \left( \vec{x} , t \right).
\end{equation}

Here ${\cal H}_0 \left( \vec{x} \right) $ is the Hamiltonian of the system without external field, and $V \left( \vec{x} , t \right)$ is the interaction of the system with external field. If there is no external field is applied,

\begin{equation}
\widehat{\cal H }_0 \left( \vec{x} \right) \Phi_n \left( \vec{x} \right) = E_n \Phi_n \left( \vec{x} \right),
\end{equation}
here $E_n$ is the eigenvalue of eigenstate $n$, and $\Phi_n \left( \vec{x} \right), n=0,1,2\dots$ are the correspond orthonormal eigenvectors.

Assume $\Psi \left( \vec{x}, t \right)$ is a superposition of $\Phi \left( \vec{x} \right)$'s
 
\begin{equation}
\label{eq:EPSIDECOMT}
\vert \bar{\Psi} \left( \vec{x},t \right) \rbrack \rangle=\sum_n \vert \Phi_n \left( \vec{x} \right) \rangle \vert \bar{C}_n \left( t \right) \rbrack.
\end{equation}

Here we use $\vert \Phi \left( \vec{x} \right) \rangle$ for states in space domain, $\vert C \left( t \right) \rbrack$ for states in time domain, and $\vert \Phi \left(\vec{x}, t\right) \rbrack \rangle$ for states in space-time domain. The difference between $\vert \rangle$ and $\vert \rbrack$ is the definition of the correspond left vector, $\langle \vec{x} \vert = \vert \vec{x} \rangle^{\dag}$ and $\lbrack t \vert = \vert t \rbrack^{T}$.

\begin{figure}
\includegraphics[width=.8500\columnwidth]{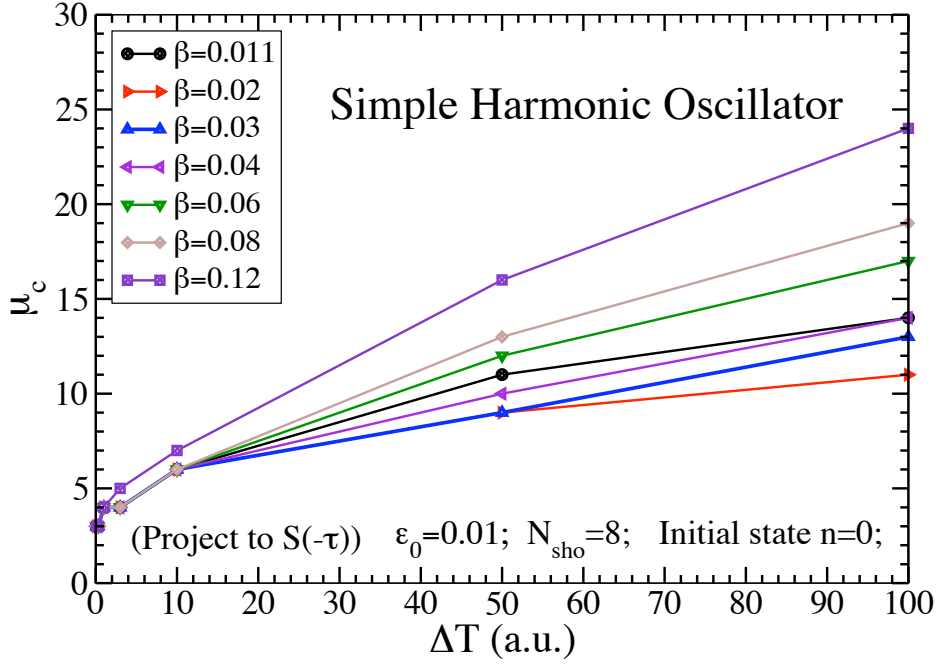}
\caption[]
{(color online) $N_{\mu}$ to $\Delta T$}
\label{cj_bt_fig11}
\end{figure}

In order to solve the TDSE, we break the time axis into finite elements and calculate the solution on each time segment. 
Without losing generality, the time nodes can be labeled as $t_j,\ j=0,1,2,3$. For simplicity, time intervals are chosen to be same length, $t_{e+1}-t_e=\Delta t$, $t_e\ \left( e=0,1,2\dots \right)$ is time node. On each time interval, $[t_{e},t_{e+1}$, define local time $\tau \in [-1,1]$,
  
\begin{equation}
\tau=B+Ct, 
\end{equation}

where

\begin{eqnarray}
\label{eq:BC}
B &\equiv& \frac{t_e + t_{e+1}} {t_e - t_{e+1}}\\
C &\equiv& \frac{2} {t_{e+1} - t_e}
\end{eqnarray}

Thus the time-dependent Schr{\"o}dinger equation Eq.~(\ref{eq:ETDSET}) and its solution Eq.~(\ref{eq:EPSIDECOMT}) can be written in terms of local time $\tau$, 
\begin{equation}
\label{eq:ETDSETAU}
\lbrack iC\hbar\frac{\partial}{\partial \tau}  - \widehat{\bar{\cal H }}^{ \left( e \right)} \left( \vec{x} , \tau \right) \rbrack 
\vert \bar{\Psi}^{ \left( e \right)} \left( \vec{x} , \tau \right) \rbrack =0,
\end{equation}

\begin{equation}
\label{eq:EPSIDECOMTAU}
\vert \bar{\Psi}^{ \left( e \right) } \left( \vec{x} , \tau \right) \rbrack \rangle=\sum_n \vert \Phi_n \left( \vec{x} \right) \rangle \vert \bar{C}^{ \left( e \right) }_n \left( \tau \right) \rbrack.
\end{equation}

Substitute Eq.~(\ref{eq:EPSIDECOMTAU}) into Eq.~(\ref{eq:ETDSETAU}) we obtain,

\begin{equation}
\sum_{n} \lbrack iC\hbar\frac{\partial}{\partial \tau}-E_n-\bar{V}^{ \left( e \right) } \left( \vec{x} , \tau \right) \rbrack \vert \Phi_n \left( \vec{x} \right) \rangle \vert \bar{C}^{ \left( e \right) }_n \left( \tau \right) \rbrack = 0. 
\end{equation}

\begin{figure}
\includegraphics[width=.8500\columnwidth]{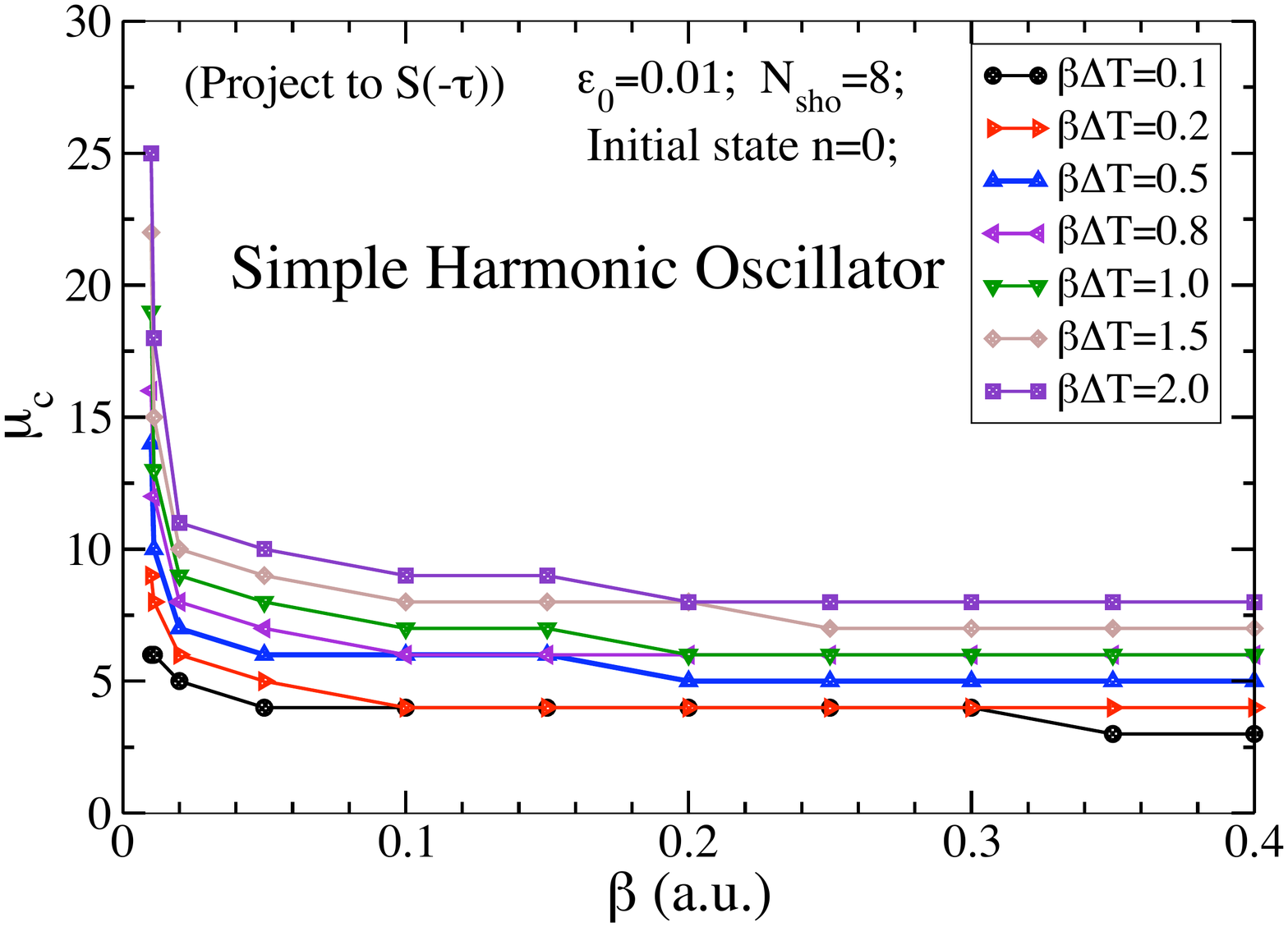}
\caption[]
{(color online) $N_{\mu}$ to $\beta \Delta T$}
\label{cj_deltaphase}
\end{figure}

%$ \lbrack \bar{C}^{ \left( e \right) }_n \left( \tau \right) \vert \langle \Phi_n \left( \vec{x} \right) \vert $

Project the above equation onto $ \langle \Phi_n \left( \vec{x} \right) \vert $,

\begin{equation}
\label{eq:ESIMUEQCNTAU}
\sum_{n} \lbrack \left( iC\hbar\frac{\partial}{\partial \tau}  - E_n \right) \delta_{n^{\prime} n} - \bar{V}^{ \left( e \right) } \left( \tau \right)_{ n^{\prime} n} \rbrack \vert \bar{C}^{ \left( e \right) }_n \left( \tau \right) \rbrack = 0.
\end{equation}
Here,

\begin{equation}
\bar{V}^{ \left( e \right)} \left( \tau \right)_{n^{\prime} n} = \langle \Phi_n \left( \vec{x} \right) \vert 
\bar{V}^{ \left( e \right)} \left( \vec{x} , \tau \right) \vert \Phi_n \left( \vec{x} \right) \rangle
\end{equation}

The wave functions must be continuous at time node $t_e$, consider Eq.~(\ref{eq:EPSIDECOMT}), the decomposition coefficients must statisfy,

\begin{equation}
\vert \bar{C}^{ \left( e \right) }_n \left( -1 \right) \rbrack = \vert \bar{C}^{ \left( e-1 \right) }_n \left( 1 \right) \rbrack.
\end{equation}

\begin{figure}
\includegraphics[width=.8500\columnwidth]{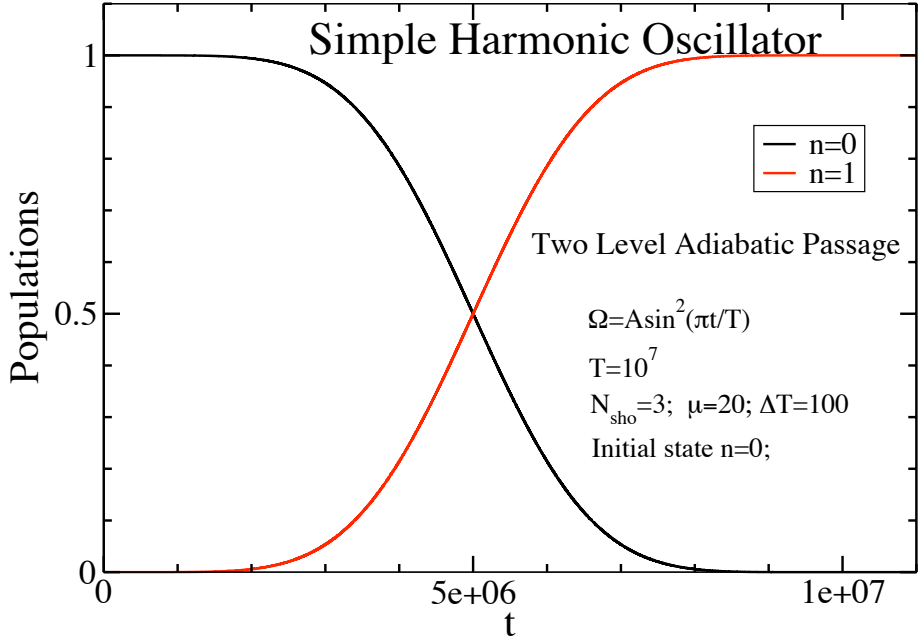}
\caption[]
{(color online) Two level adiabatic passage with $2$ harmonic states. Here $\Omega \left( t \right) =A_0 \sin^2 \frac{\pi t}{T}$.}
\label{cj_sho2lev2s}
\end{figure}

The coefficients in Eq.~(\ref{eq:ETDSETAU}) can be expressed as

\begin{equation}
\label{eq:ECENTAUORI}
\vert \bar{C}^{ \left( e \right) }_n \left( \tau \right) \rbrack = \vert f^{ \left( e \right) }_n \left( \tau \right) \rbrack
+\vert \bar{C}^{ \left( e \right) }_n \left( -1 \right) \rbrack,
\end{equation}
where $\vert f^{ \left( e \right) }_n \left( \tau \right) \rbrack $ satisfies,

\begin{equation}
\vert f^{ \left( e \right) }_n \left(-1\right) \rbrack = 0.
\end{equation}

The local function $f^{\left(e\right)}_n \left(\tau\right)$ is expanded in time basis $S_{\mu}\left(\tau\right)$ as

\begin{equation}
\vert f^{\left(e\right)}_n \left(\tau\right) \rbrack = \sum^{N_{\mu}-1}_{\mu=0} \vert S_{\mu}\left(\tau\right) \rbrack \bar{B}^{\left(e\right)}_{\mu n}.
\end{equation}

In the present work, the time basis $S_{\mu}\left(\tau\right)$ is defined from Chebyshev Polynomials \cite{JHESTHAVEN07B},

\begin{equation}
S_{\mu}\left(\tau\right) = \int^{\tau}_{-1}T_{\mu}\left(\tau\right) \mathrm{d} \tau,
\end{equation}
where $T_{\mu}\left(\tau\right)$ is the first kind Chebyshev Polynomial.

Now Eq.~(\ref{eq:ECENTAUORI}) can be rewrite as,

\begin{eqnarray}
\label{eq:ECENTAU}
\vert \bar{C}^{\left(e\right)}_n \left(\tau\right) \rbrack &=& \sum^{N_{\mu}-1}_{\mu=0} \vert S_{\mu}\left(\tau\right) \rbrack \bar{B}^{\left(e\right)}_{\mu n}+\vert \bar{C}^{\left(e\right)}_n \left(-1\right) \rbrack \nonumber\\
&=& \sum^{N_{\mu}-1}_{\mu=0} \vert S_{\mu}\left(\tau\right) \rbrack \bar{B}^{\left(e\right)}_{\mu n}
+\vert \bar{C}^{\left(e-1\right)}_n \left(1\right) \rbrack.
\end{eqnarray}

Substitute Eq.~(\ref{eq:ECENTAU}) into Eq.~(\ref{eq:ESIMUEQCNTAU}),

\begin{eqnarray}
\label{eq:ESIMUEQB}
& &\sum_{\mu n} \lbrack \left(iC\hbar\frac{\partial}{\partial \tau}
- E_n\right) \delta_{n^{\prime} n} - \bar{V}^{\left(e\right)} \left(\tau\right)_{n^{\prime} n} \rbrack 
\vert S_{\mu} \left(\tau\right) \rbrack \bar{B}^{\left(e\right)}_{\mu n} \nonumber\\
&=& \sum_n \lbrack E_n \delta_{n^{\prime} n} + \bar{V}^{\left(e\right)}_{n^{\prime} n} \left(\tau\right) \rbrack \bar{C}^{\left(e-1\right)}_n \left(1\right).
\end{eqnarray}

\begin{figure}
\includegraphics[width=.8500\columnwidth]{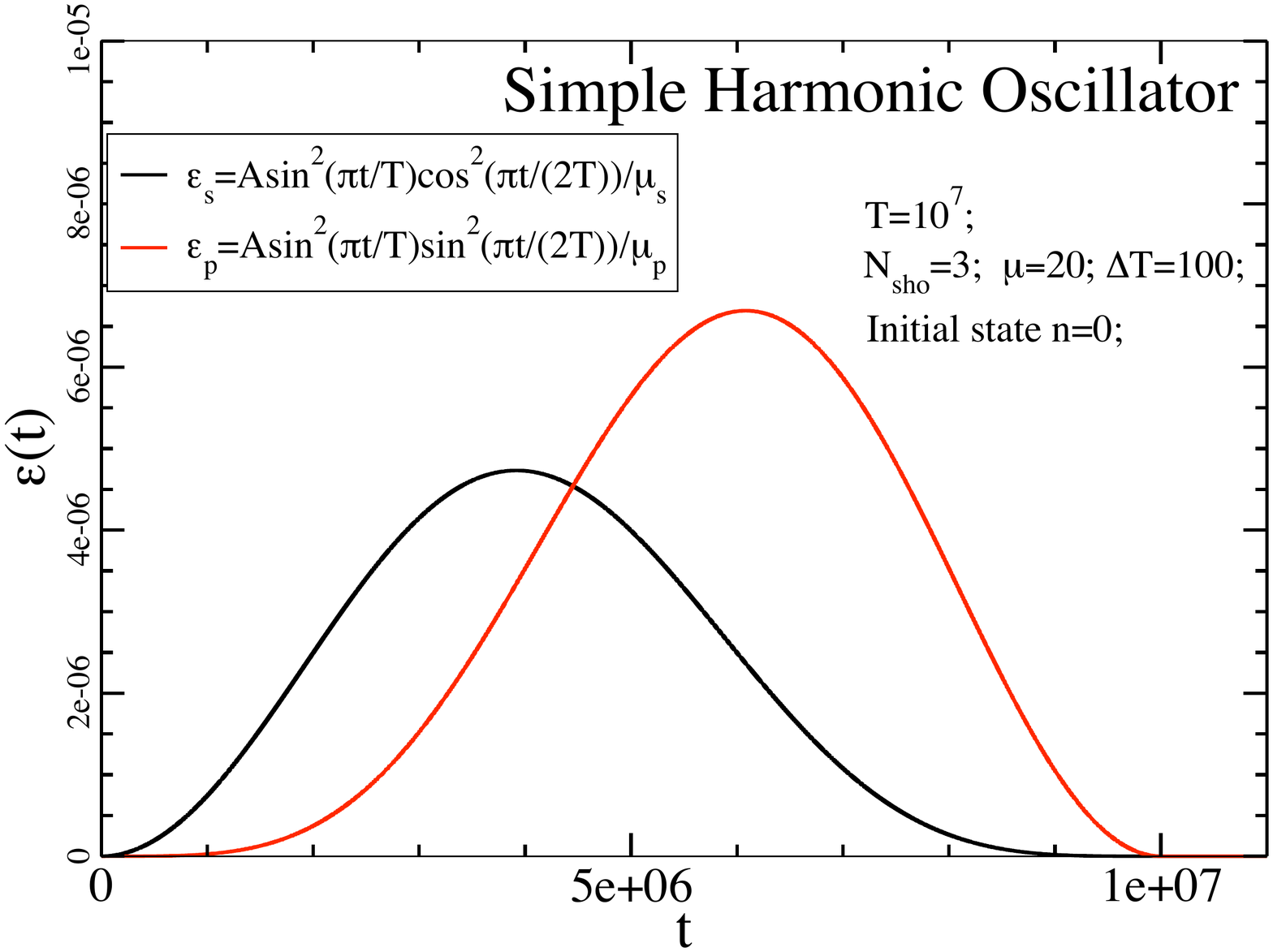}
\caption[]
{(color online) Laser pulse in STIRAP for a simple harmonic oscillator in laser field. Here $\Omega_s \left( t \right) =A_0 \cos^2\frac{\pi t}{2T} \sin^2 \frac{\pi t}{T}$,
$\Omega_p \left( t \right) =A_0 \sin^2\frac{\pi t}{2T} \sin^2 \frac{\pi t}{T}$.}
\label{cj_sho3lsinxov2laser}
\end{figure}

After projecting Eq.~(\ref{eq:ESIMUEQB}) onto $\lbrack S_{\mu^{\prime} } \left(\tau\right) \vert \omega \left(\tau\right)$, we obtain, 

\begin{eqnarray}
\label{eq:ESIMU}
& & \sum_{\mu n} \lbrack \left(iCd_{\mu^{\prime} \mu}
- E_n O_{\mu^{\prime} \mu} \right) \delta_{n^{\prime} n} - \bar{V}^{\left(e\right)} \left(\tau\right)_{\left(\mu n \right)^{\prime} \left(\mu n\right)} \rbrack 
\bar{B}^{\left(e\right)}_{\mu n}\nonumber\\
&=& \sum_n \lbrack E_n g_{\mu^{\prime}}\delta_{n^{\prime} n} + \bar{V}^{\left(e\right)}_{\left(\mu n \right)^{\prime} n} \rbrack \bar{C}^{\left(e-1\right)}_n \left(1\right)
\end{eqnarray}

Here $\omega \left(\tau\right)$ is Chebyshev weight function,
\begin{equation}
\label{eq:EWEIGHT}
\omega \left(\tau\right) = \left(1 - {\tau}^2\right)^{\frac{1}{2}},
\end{equation}
and

\begin{eqnarray}
\label{eq:EDELEM}
d_{ \mu^{ \prime } \mu} &\equiv& \lbrack S_{ \mu^{ \prime } } \left(\tau \right) \vert \omega \left( \tau \right) \vert T_{ \mu } \left( \tau \right) \rbrack\\
\label{eq:EOMEGAELEM}
O_{ \mu^{ \prime } \mu} &\equiv& \lbrack S_{ \mu^{ \prime } } \left(\tau \right) \vert \omega \left( \tau \right) \vert S_{ \mu } \left( \tau \right) \rbrack\\
\label{eq:EGELEM}
g_{\mu^{\prime}} &\equiv& \lbrack S_{ \mu^{\prime}}\left(\tau\right) 
\vert \omega \left( \tau \right) \rbrack\\
\label{eq:EVBARELEM}
\bar{V}^{\left(e\right)}_{ \left(\mu n\right)^{ \prime } \left(\mu n\right) } &\equiv& \lbrack S_{ \mu^{ \prime } } \left(\tau\right) \omega \left( \tau \right) \vert \bar{V}^{\left(e\right)}_{ n^{\prime} n } \left( \tau \right) 
\vert S_{ \mu } \left( \tau \right) \rbrack
\end{eqnarray}

The Finite-Element Space-Time algorithm is implemented in the following steps:

1. Choose the number of space bases, $N$, and the number of time bases, $N_{\mu}$.

2. Choose time step $\delta t$. Start calculating from $e=0, t_e=t_0=0$.

3. At the $e^{\it th}$ step

\begin{itemize}
\item calculate $t_{e+1}=t_e + \delta t$, and $B,C$ by using Eq.~(\ref{eq:BC}).
\item for $\mu^{\prime},\mu=0,1,2\dots N_{\mu-1}$, calculate elements in Eq.~(\ref{eq:ESIMU}) by using Eq.~(\ref{eq:EDELEM}) to Eq.~(\ref{eq:EVBARELEM}).
\item solve simultaneous equation Eq.~(\ref{eq:ESIMU}) to obtain $B^{\left(e\right)}_{\mu n}$.
\item calculate $\vert C_n \left(t_{e+1}\right) \rbrack$ and $\Psi \left(t_{e+1}\right)$.
\end{itemize}

4. Repeat (3) until time $t$.

\begin{figure}
\includegraphics[width=.8500\columnwidth]{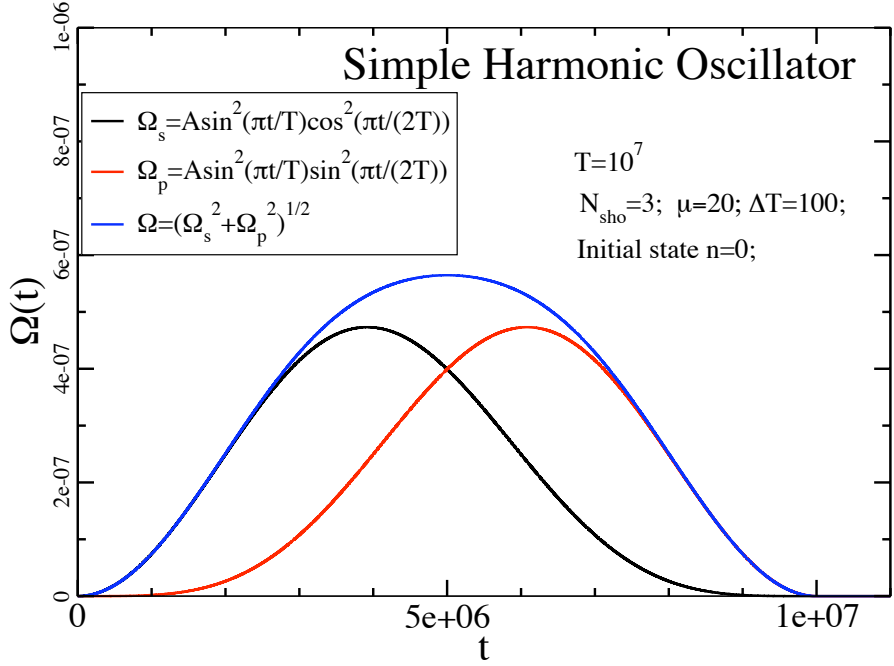}
\caption[]
{(color online) Rabi frequencies in STIRAP procession for simple harmonic oscillator in laser field. Here $\Omega_s \left( t \right) =A_0 \cos^2\frac{\pi t}{2T} \sin^2 \frac{\pi t}{T}$, $\Omega_p \left( t \right) =A_0 \sin^2\frac{\pi t}{2T}
\sin^2 \frac{\pi t}{T}$.}
\label{cj_sho3lsinxov2omega}
\end{figure}

\section{Model System and Adiabatic Passage}
\subsection{Model System}

The Model System used in the presented work is a simple harmonic oscillator with an external laser field applied on the oscillator. The Hamiltonian of a free harmonic oscillator is,

\begin{equation}
\label{eq:HAMILT}
{\cal H}={\cal H}_0+V.
\end{equation}

Here ${\cal H}_0$ is the Hamiltonian for simple harmonic oscillator without external field,   

\begin{equation}
{\cal H}_0=-\frac{\hbar^2 \nabla^2}{2m}+\frac{m\omega^2 x^2}{2}.
\end{equation}

The eigenvalues of $\cal H$ are
\begin{equation}
E=\left(\frac{1}{2}+n\right)\hbar \omega,
\end{equation}

here $n$ is the principle number and $n=0,1,2\cdots$.

\begin{figure}
\includegraphics[width=.8500\columnwidth]{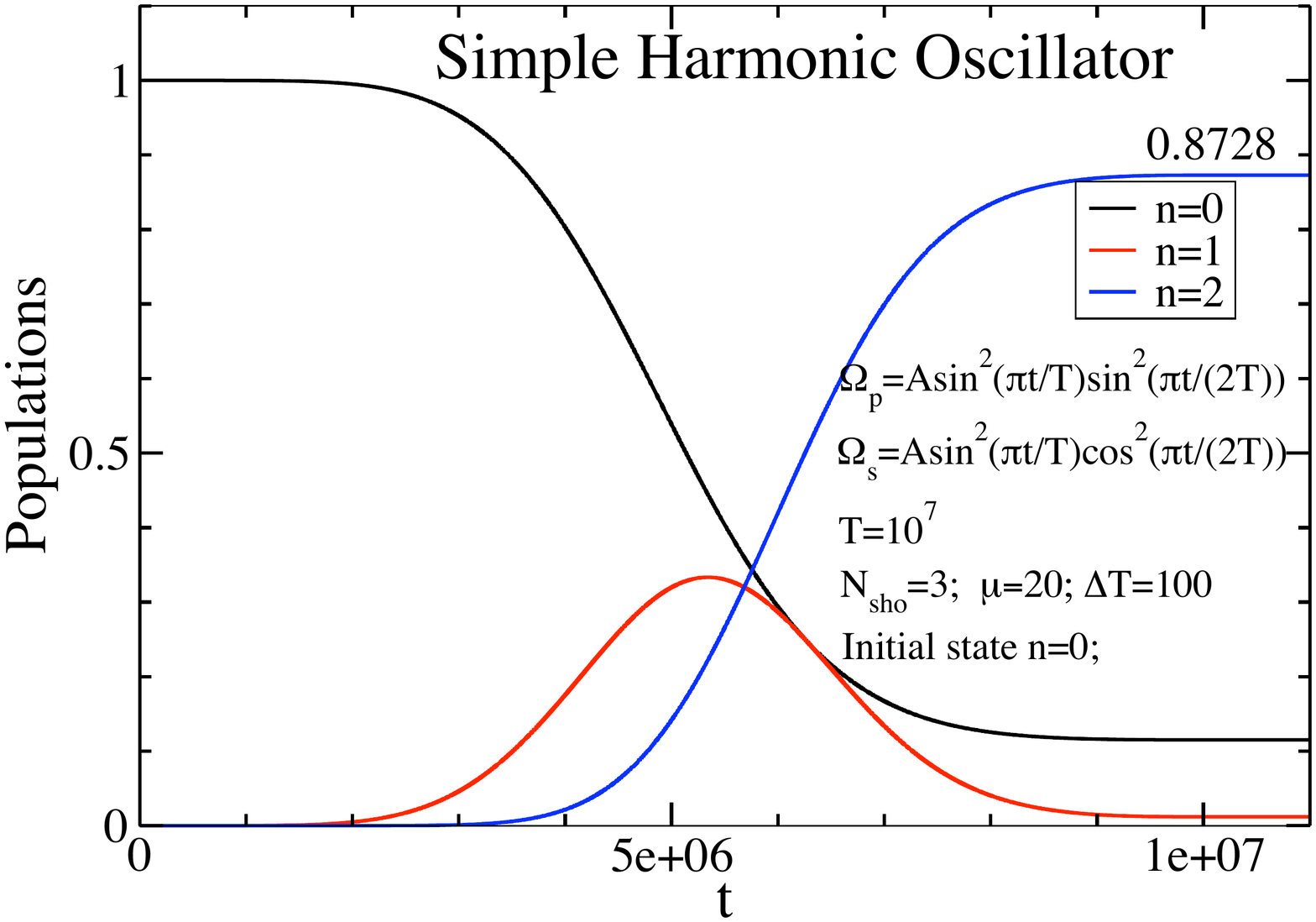}
\caption[]
{(color online) Population transfer in STIRAP procession for simple harmonic oscillator in laser field. Here $\Omega_s \left( t \right) =A_0 \cos^2\frac{\pi t}{2T} \sin^2 \frac{\pi t}{T}$,
$\Omega_p \left( t \right) =A_0 \sin^2\frac{\pi t}{2T} \sin^2 \frac{\pi t}{T}$.
}
\label{cj_sho3lsinxov2prb}
\end{figure}

The $V$ in Eq.~(\ref{eq:HAMILT}) is the dipole interaction and can be expressed as
\begin{equation}
V=-\overrightarrow{r}\left(t\right)\cdot\overrightarrow{f}\left(t\right)
\end{equation}

External laser field can be expressed as,

\begin{equation}
\overrightarrow{f}\left(t\right)=\overrightarrow{\varepsilon} \left(t\right)\cos\left(\omega t\right)
\end{equation}

for single pulse (used in two-level adiabatic passage) and 

\begin{eqnarray}
\overrightarrow{f}\left(t\right)&=&\overrightarrow{\varepsilon_p} \left(t\right)\cos\left(\omega_p t+\varphi_p\right)+\overrightarrow{\varepsilon_s} \left(t\right)\cos\left(\omega_s t + \varphi_s\right)\nonumber\\
\label{eq:E3LASERFIELD}
&=& \left( \overrightarrow{\varepsilon_p} \left(t\right) +\overrightarrow{\varepsilon_s} \left(t\right) \right) \cos\left(\omega t\right)
\end{eqnarray}
for two-pulse (used in STIRAP). Here $\omega$ is excitation frequency and $\varepsilon$ is pulse contour. In the case for coherent control of a Simple Harmonic Oscillator interaction with laser, $\omega_s=\omega_p=\omega$. For simplicity, we choose $\varphi_p=\varphi_s=0$.

\subsection{Two-Level Adiabatic Passage for Simple Harmonic Oscillator}

We first considered the adiabatic passage for one-dimension simple harmonic oscillator with Hamiltonian ${\cal H}_0$ having only two eigenstates, $\psi_a$ and $\psi_b$. For simplicity, we take $\psi_a=\psi_0$ and $\psi_b=\psi_1$. From \cite{DTANNOR07B}, if $\omega_0$ is defined as $\omega_0=\frac{E_b-E_a}{\hbar}$, the detuning is defined as $\Delta=\omega-\omega_0$. The Rabi frequency is

\begin{equation}
\Omega=\sqrt{{\Delta}^2+{\left(\frac{\mu \varepsilon}{\hbar}\right)}^2}.
\end{equation}

\begin{figure}
\includegraphics[width=.8500\columnwidth]{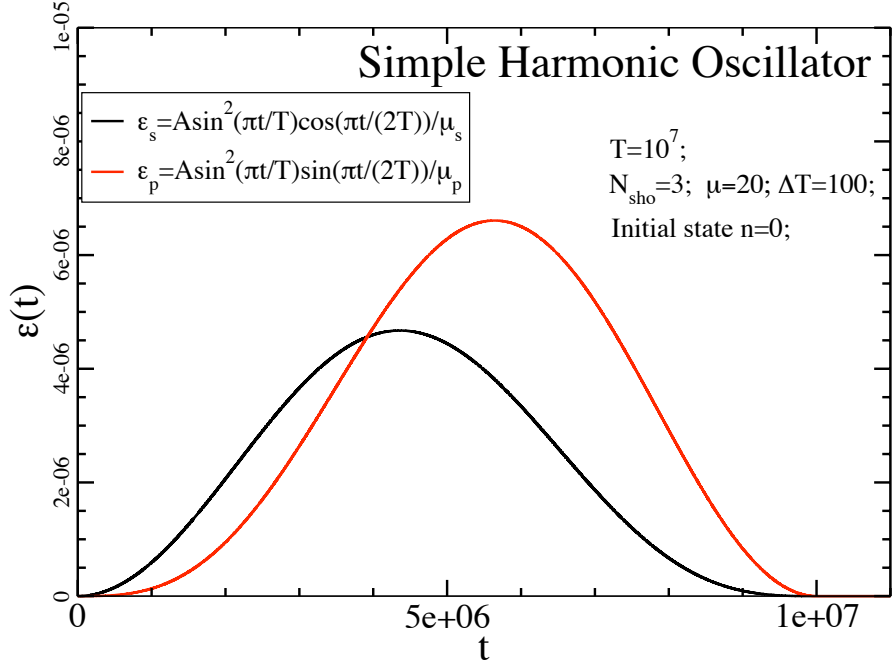}
\caption[]
{(color online) Laser pulse in STIRAP for a simple harmonic oscillator in laser field. Here $\Omega_s \left( t \right) =A_0 \cos\frac{\pi t}{2T} \sin^2 \frac{\pi t}{T}$,
$\Omega_p \left( t \right) =A_0 \sin\frac{\pi t}{2T} \sin^2 \frac{\pi t}{T}$.}
\label{cj_sho3llaser}
\end{figure}

In resonant excitation process, $\Delta=0$, then the Rabi frequency is

\begin{equation}
\Omega=\frac{\mu \varepsilon}{\hbar}.
\end{equation}
For one-dimension simple harmonic oscillator, 

\begin{equation}
\varepsilon= \varepsilon_0 \sin^2\frac{\pi t}{T}.
\end{equation}

\begin{figure}
\includegraphics[width=.8500\columnwidth]{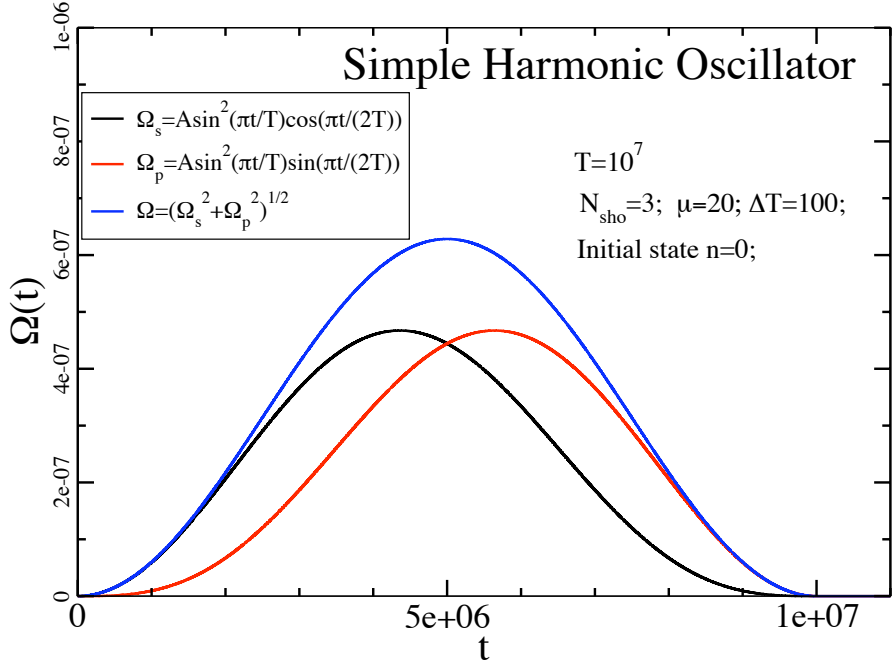}
\caption[]
{(color online) Rabi frequencies in STIRAP procession for simple harmonic oscillator in laser field. Here $\Omega_s \left( t \right) =A_0 \cos\frac{\pi t}{2T} \sin^2 \frac{\pi t}{T}$,
$\Omega_p \left( t \right) =A_0 \sin\frac{\pi t}{2T} \sin^2 \frac{\pi t}{T}$.}
\label{cj_sho3lomega}
\end{figure}

Then $\Omega$ can be written as,

\begin{equation}
\Omega \left( t \right) = A_0 \sin^2\frac{\pi t}{T}.
\end{equation}

In the above equation, 
\begin{eqnarray}
\mu &=& \mu_{ab}=\mu_{ba}\\
&=& \langle a \vert r \vert b \rangle
= \langle b \vert r \vert a \rangle.
\end{eqnarray}

From the pulse area theorem in section 15.1 in \cite{DTANNOR07B}, Eq.~(15.17), for resonant excitation, if the pulse duration and Rabi frequency satisfy the following relation,
\begin{equation}
\label{eq:pulseareatheorem}
\int_0^t \frac{\Omega\left( t^{\prime} \right)\mathrm {d} t^{\prime}}{2}=\frac{\pi}{2},
\end{equation}
the population is fully transferred to the upper state at time $t$. That means if the pulse length is $T$, and population is expected to be fully transferred to the upper state, the amplitude factor $A_0$ and $\varepsilon_0$ should be,

\begin{equation}
A_0=\frac{2\pi}{T}
\end{equation}

\begin{equation}
\varepsilon_0=\frac{2\pi \hbar}{\mu T}
\end{equation}

\subsection{Three-level Stimulated Raman Adiabatic Passage (STIRAP) for Simple Harmonic Oscillator}

The STIRAP procession involves at least three states. We here consider the case with only three energy eigenstates, the ground state $\vert 0 \rangle$, the first excited state $\vert 1 \rangle$ and the second excited state $\vert 2 \rangle$. For a simple harmonic oscillator, the STIRAP procession is ladder type. The population transfers from $\vert 0 \rangle$ to $\vert 1 \rangle$ and then to $\vert 2 \rangle$, where $\vert 1 \rangle$ is intermediate state.

The two pulses are in counterintuitive order. For coherent control, the pump pulse should be close to the resonant of $\vert 0 \rangle \leftrightarrow \vert 1 \rangle$ transition and the Stokes pulse should be close to the resonant of  $\vert 1 \rangle \leftrightarrow \vert 2 \rangle$ transition. Thus we set $\Delta_p=\Delta_s=\Delta=0$ and have,

\begin{eqnarray}
\Omega_p&=&\frac{\mu_p \varepsilon_p}{\hbar},\\
\Omega_s&=&\frac{\mu_s \varepsilon_s}{\hbar}.
\end{eqnarray}

Here $\mu_s$ and $\mu_p$ are 

\begin{eqnarray}
\mu_s &=& \langle 1 \vert r \vert 2 \rangle,\\
\mu_p &=& \langle 0 \vert r \vert 1 \rangle.
\end{eqnarray}

The Rabi frequency is

\begin{equation}
\label{eq:ERABI3}
\Omega=\sqrt{{{\left(\frac{\mu_s \varepsilon_s}{\hbar}\right)}^2 + {\left(\frac{\mu_p \varepsilon_p}{\hbar}\right)}^2}}.
\end{equation}

\begin{figure}
\includegraphics[width=.8500\columnwidth]{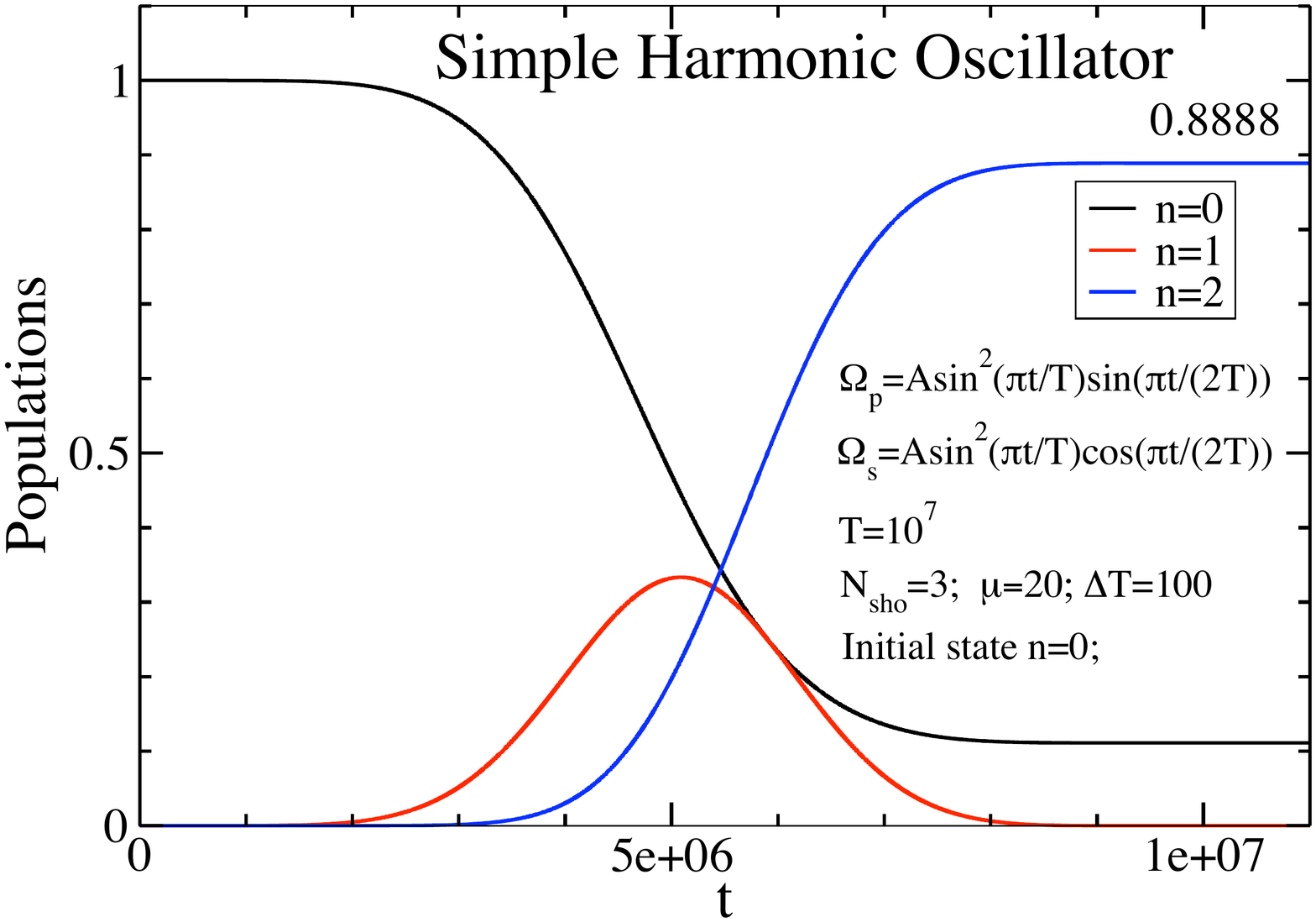}
\caption[]
{(color online) Population transfer in STIRAP procession for simple harmonic oscillator in laser field. Here $\Omega_s \left( t \right) =A_0 \cos\frac{\pi t}{2T} \sin^2 \frac{\pi t}{T}$,
$\Omega_p \left( t \right) =A_0 \sin\frac{\pi t}{2T} \sin^2 \frac{\pi t}{T}$.
}
\label{cj_sho3lprb}
\end{figure}

We write the contour of two laser pulses as

\begin{equation}
\label{eq:ELASERAMPSTOKES}
\varepsilon_s = \varepsilon_{s0} s \left( t \right),% \sin^2\frac{\pi t}{T},
\end{equation}
\begin{equation}
\label{eq:ELASERAMPPUMP}
\varepsilon_p = \varepsilon_{p0} p \left( t \right),% \sin^2\frac{\pi t}{T}.
\end{equation}

Thus we obtain
\begin{equation}
\label{eq:ERABISTOKES1}
\Omega_s \left( t \right)=\left( \frac{\mu_s \varepsilon_{s0} }{\hbar} \right) s \left( t \right),% \sin^2\frac{\pi t}{T}
\end{equation}

\begin{equation}
\label{eq:ERABIPUMP1}
\Omega_p \left( t \right)=\left( \frac{\mu_p \varepsilon_{p0} }{\hbar} \right) p \left( t \right). % \sin^2\frac{\pi t}{T}
\end{equation}

we assume
\begin{equation}
\label{eq:EA0}
\frac{\mu_s \varepsilon_{s0} } {\hbar} = \frac{ \mu_p \varepsilon_{p0} } {\hbar} = A_0. 
\end{equation}

Then $\Omega_s, \Omega_p$ can be written as
\begin{equation}
\label{eq:ERABISTOKES2}
\Omega_s \left( t \right) =A_0 s \left( t \right),% \sin^2 \frac{\pi t}{T},
\end{equation}

\begin{equation}
\label{eq:ERABIPUMP2}
\Omega_p \left( t \right) =A_0 p \left( t \right).% \sin^2 \frac{\pi t}{T}.
\end{equation}

Thus Rabi frequency $\Omega$ is
\begin{eqnarray}
\label{eq:ERABI4}
\Omega \left( t \right)
&=& \left( \Omega_s^2 \left( t \right)+ \Omega_p^2 \left( t \right) \right)^{\frac{1}{2}}\nonumber\\
&=& A_0 \left( s^2 \left( t \right) + p^2 \left( t \right) \right)^{\frac{1}{2}}.% \sin^2\frac{\pi t}{T}. 
\end{eqnarray}

Since Rabi frequency $\Omega$ is related to the final state and the initial state, we assume for complete population transfer, the pulse area theorem still holds, thus,

\begin{eqnarray}
\label{eq:pulseareatheorem2}
& &\int_0^t \frac{\Omega\left( t^{\prime} \right)\mathrm {d} t^{\prime}}{2} \nonumber\\
&=& \int_0^t \frac{A_0 \left(s^2\left(t^{\prime}\right)+p^2\left(t^{\prime}\right)\right)^{\frac{1}{2}}
%\sin^2\frac{\pi t^{\prime}}{T}
\mathrm {d}t^{\prime}}{2}
= \frac {\pi} {2}.
\end{eqnarray}

%(Note: I tried to prove the above equation analytically, but so far I failed to do that).

So that, if at time $t=T$, the population is fully transferred to the second excited state, $A_0$ should be

\begin{equation}
\label{eq:E3A0}
A_0=\frac{\pi} {\int_0^T \left( s^2 \left( t^{\prime} \right) + p^2 \left( t^{\prime} \right) \right)^{\frac{1}{2}}
% \sin^2\frac{\pi t^{\prime}}{T}
\mathrm {d} t^{\prime}}
\end{equation}

Once $s(t)$ and $p(t)$ are chosen,  $A_0$ can be calculated according to Eq.~(\ref{eq:E3A0}), then $\varepsilon_{s0}$ and $\varepsilon_{p0}$ can be obtained through

\begin{equation}
\varepsilon_{s0}=\frac{A_0 \hbar}{\mu_s},
\end{equation}

\begin{equation}
\varepsilon_{p0}=\frac{A_0 \hbar}{\mu_p}.
\end{equation}

\section{calculation results}
%section IV

In our calculation, we used the same $k$ and $m$ values as that used by Lauvergnat et al, \cite{DLAUVERGNATJCP126}, in section
V entitled "Forced harmonic Oscillator" under eq. 5.2. 

\begin{equation}
k=1, m=10000.
\end{equation}

\begin{figure}
\includegraphics[width=.8500\columnwidth]{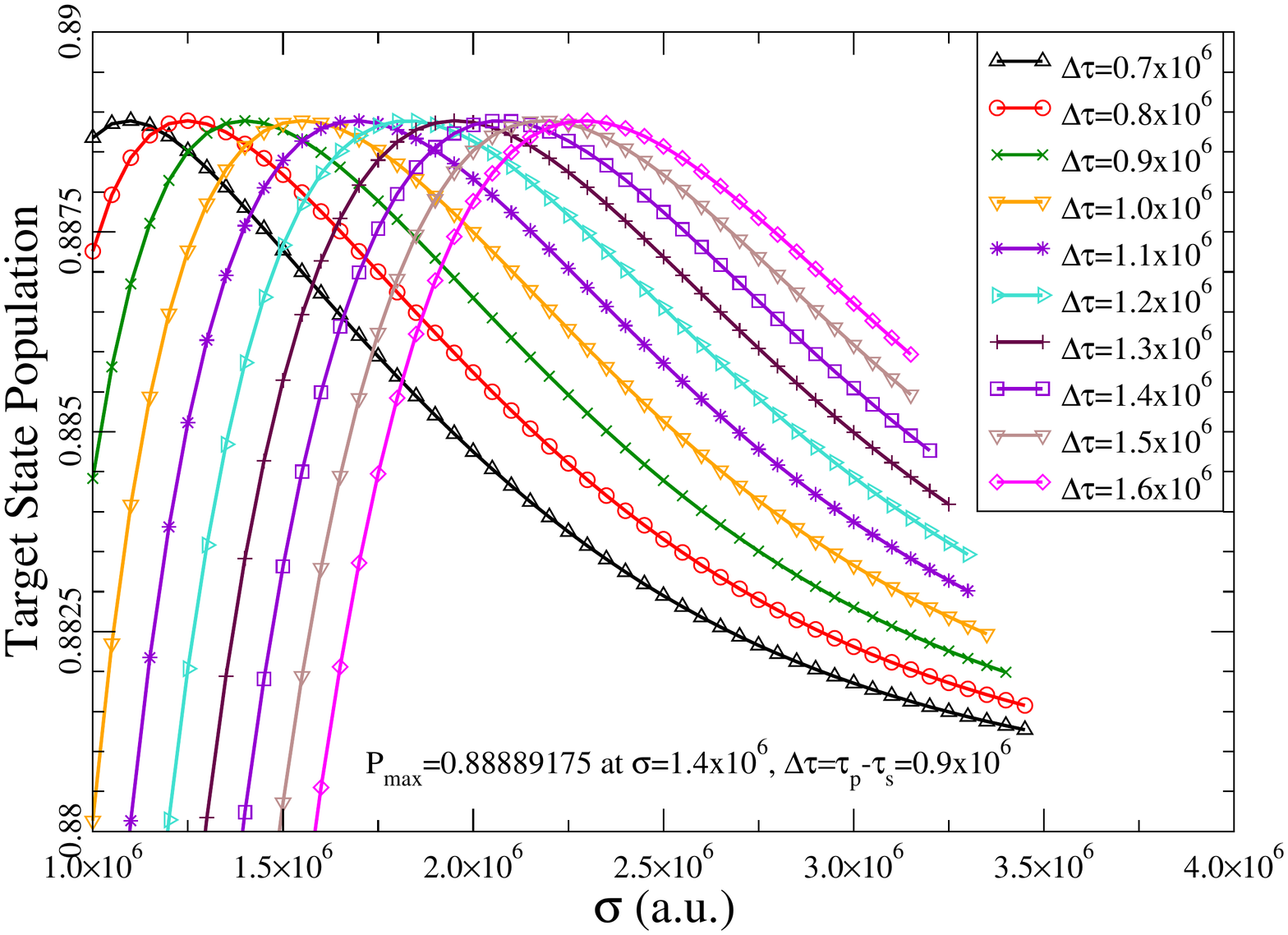}
\caption[]
{(color online) Target state population vs. the variance of Gauss pulse for defferent pulse delay in STIRAP process for a simple harmonic oscillator in Gauss pulse laser field. Here $\Omega_s \left( t \right) =A_0 \frac{1}{\sigma \sqrt{2 \pi}}e^{-(t-\tau_s)^2/(2\sigma^2)}$,$\Omega_p \left( t \right) =A_0 \frac{1}{\sigma \sqrt{2 \pi}}e^{-(t-\tau_p)^2/(2\sigma^2)}.$}
\label{cj_sho3lgausssumprb}
\end{figure}

The oscillator frequency is $\omega=\sqrt{\frac{k}{m}}=0.01 a.u. $\ The correspond resonant external laser field we applied on it has a period of $T_{res}=\frac{2 \pi }{\omega}=200 \pi a.u. = 4.8 \pi \times 10^{-15} sec = 4.8 \pi \times 10^{-3} ps \approx 15 \times 10^{-3} ps$. The wavelength of the laser is $\lambda = 452.4 nm$. In our calculation, we chose pulse duration to be to be $1000 \sim 8000$ times of the laser period. This ratio is close to that used by Sarkar et al \cite{CSARKARPRA78}.

\subsection{Verification of the Finite-Element Space-Time Algorithm}

To verify our algorithm, we use a laser pulse $f\left(t\right)$ with periodic frequency $\beta$. We vary $\beta$ in a range to analyze the relations of the parameters. 
The external laser is 

\begin{eqnarray}
f \left( t \right) &=& \varepsilon \left( t \right)\cos \left(\beta t \right)\nonumber\\
&=& \varepsilon_0 \sin^2 \frac{\pi t }{T} \cos  \left( \beta t \right).
\end{eqnarray}

Our calculation result shows that the norm of the wave function is close to $1$ with the error less than $10^{-10}$. 

If the amplitude of external field is fixed, in order to obtain the accuracy, with the error of norm less than $10^{-10}$, the number of time bases is related to the time step $\Delta T$, shown in Fig.~(\ref{cj_bt_fig11}). 

The relation of the required number of time bases and $\beta \Delta T$ is shown in Fig.~(\ref{cj_deltaphase}).

\begin{figure}
\includegraphics[width=.8500\columnwidth]{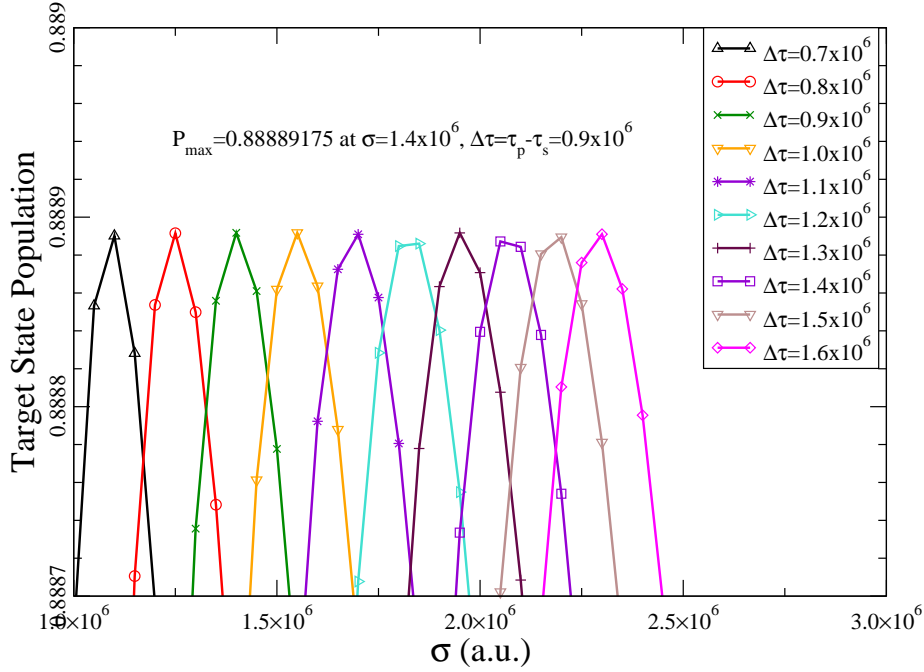}
\caption[]
{(color online) Target state population vs. the variance of Gauss pulse for defferent pulse delay in STIRAP process for a simple harmonic oscillator in Gauss pulse laser field. Only the neighbor part of the maximum populations is shown. The maximum value shown in the figure is from our calculation. Due to the random chosen of pulse delay $\Delta\tau$ and pulse width $\sigma$, the value of maximum population $0.88889175$ and the corresponding pulse delay $\Delta\tau=0.9 \times 10^6$, the pulse width $1.4 \times 10^6$ may not be correct and accurate.}
\label{cj_sho3lgausssumprbpart}
\end{figure}

\subsection{Two Level Adiabatic Passage}

In a Simple Harmonic Oscillator system with only ground state and the first excited state, if the external laser has only one pulse, the population is fully transfered to the excited state, see Fig.~(\ref{cj_sho2lev2s}).

\subsection{Stimulated Raman Adiabatic Passage}

We calculate the STIRAP process for a simple harmonic oscillator system with ground state and the first two excited states. We set $s \left(t\right)$ and $p\left(t\right)$ to be

\begin{eqnarray}
\label{eq:Est}
s \left( t \right) &=& \cos^2\frac{\pi t}{2T}\sin^2 \frac{\pi t}{T},\\
\label{eq:Esp}
p \left( t \right) &=& \sin^2\frac{\pi t}{2T}\sin^2 \frac{\pi t}{T}. 
\end{eqnarray}

Thus,

\begin{eqnarray}
\Omega_s \left( t \right) &=&A_0 \cos^2\frac{\pi t}{2T} \sin^2 \frac{\pi t}{T},\\
\Omega_p \left( t \right) &=&A_0 \sin^2\frac{\pi t}{2T} \sin^2 \frac{\pi t}{T}.
\end{eqnarray}

These results are shown in Fig.~(\ref{cj_sho3lsinxov2laser}) for external laser pulse, Fig.~(\ref{cj_sho3lsinxov2omega}) for Rabi frequency, and Fig.~(\ref{cj_sho3lsinxov2prb}) for population transfer.

\begin{figure}
\includegraphics[width=.8500\columnwidth]{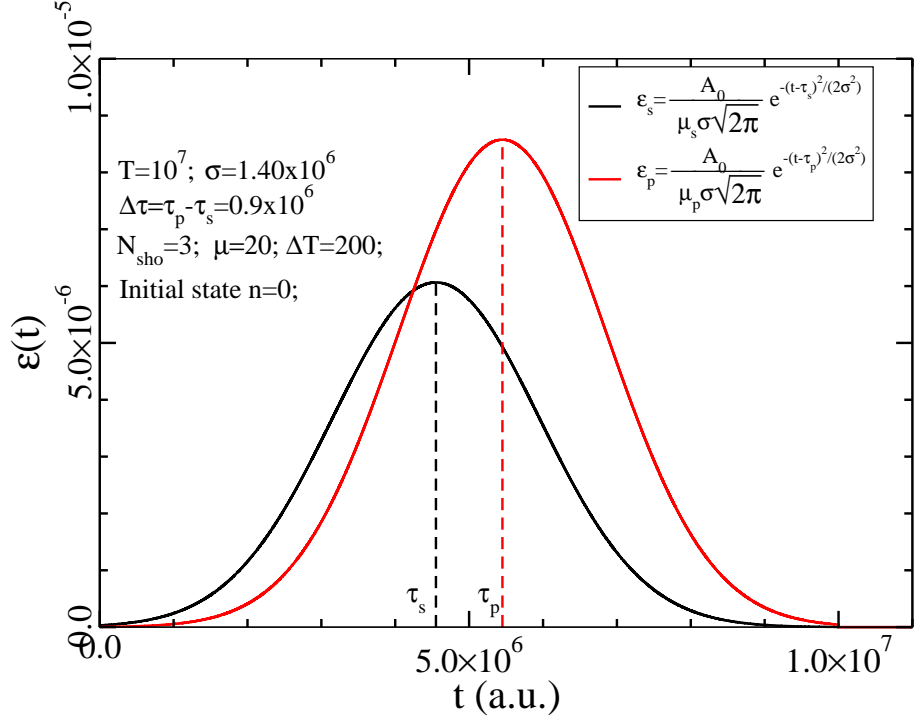}
\caption[]
{(color online) Laser pulse in STIRAP for a simple harmonic oscillator in laser field. Here $\Omega_s \left( t \right) =A_0 \frac{1}{\sigma \sqrt{2 \pi}}e^{-(t-\tau_s)^2/(2\sigma^2)}$,$\Omega_p \left( t \right) =A_0 \frac{1}{\sigma \sqrt{2 \pi}}e^{-(t-\tau_p)^2/(2\sigma^2)}.$}
\label{cj_sho3lgausslaser}
\end{figure}

We also set $s\left(t\right)$ and $p\left(t\right)$ to be

\begin{eqnarray}
\label{eq:Est2}
s \left( t \right) &=& \cos \frac{\pi t}{2T}\sin^2 \frac{\pi t}{T},\\
\label{eq:Esp2}
p \left( t \right) &=& \sin \frac{\pi t}{2T}\sin^2 \frac{\pi t}{T},
\end{eqnarray}

and calculated the population transfers.

In this case, $\Omega_s \left( t \right)$ and $\Omega_p \left( t \right)$ are

\begin{eqnarray}
\Omega_s \left( t \right) &=& A_0 \cos\frac{\pi t}{2T} \sin^2 \frac{\pi t}{T},\\
\Omega_p \left( t \right) &=& A_0 \sin\frac{\pi t}{2T} \sin^2 \frac{\pi t}{T}.
\end{eqnarray}

The results are shown in Fig.~(\ref{cj_sho3llaser}) for external laser pulse, Fig.~(\ref{cj_sho3lomega}) for Rabi frequency, and Fig.~(\ref{cj_sho3lprb}) for population transfer.

\begin{figure}
\includegraphics[width=.8500\columnwidth]{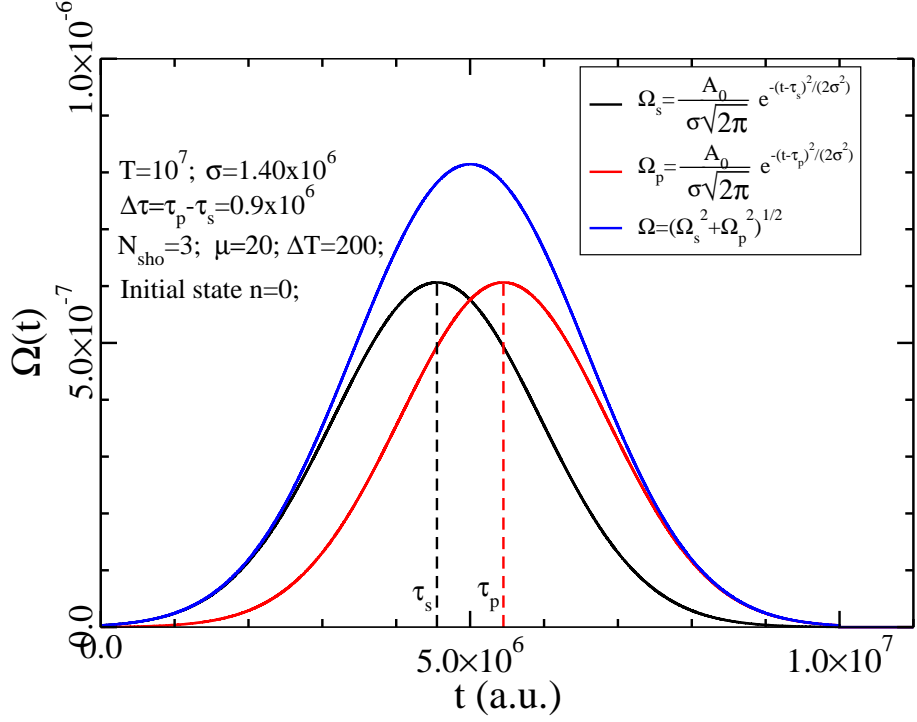}
\caption[]
{(color online) Rabi frequencies in STIRAP procession for simple harmonic oscillator in laser field. Here $\Omega_s \left( t \right) =A_0 \frac{1}{\sigma \sqrt{2 \pi}}e^{-(t-\tau_s)^2/(2\sigma^2)}$,
$\Omega_p \left( t \right) =A_0 \frac{1}{\sigma \sqrt{2 \pi}}e^{-(t-\tau_p)^2/(2\sigma^2)}.$}
\label{cj_sho3lgaussomega}
\end{figure}

For the two sets of $s\left(t\right)$ and $p\left(t\right)$ we used, the calculation results shown the population transfer to the second excited state are more than $88\%$. 

%(Note: When I did integration, I used Eq.~(7) in \cite{RBAERPRA62} to simplify programming. The error of this equation is about $0.1\%$, which is large (You told us and I verified it). I will go on programming with Eq.~(18) in \cite{RBAERPRA62}. Eq.~(18) is more accurate, I had used Eq.~(18) for $H$ atom and verified that. So at this stage, I can NOT make a conclusion which of the two pulse sets I used can mostly transfer the population to the second excited state. )

\subsection{Stimulated Raman Adiabatic Passage with Gauss-Shape Laser Field}
\label{subsec:PGAUSS}

\begin{figure}
\includegraphics[width=.8500\columnwidth]{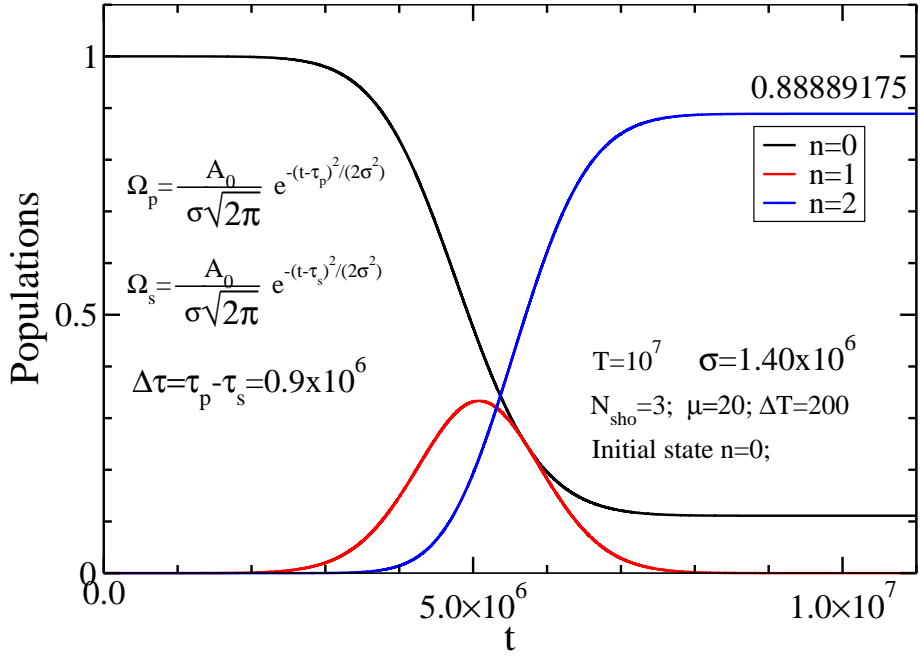}
\caption[]
{(color online) Population transfer in STIRAP procession for simple harmonic oscillator in laser field. Here $\Omega_s \left( t \right) =A_0 \frac{1}{\sigma \sqrt{2 \pi}}e^{-(t-\tau_s)^2/(2\sigma^2)}$,
$\Omega_p \left( t \right) =A_0 \frac{1}{\sigma \sqrt{2 \pi}}e^{-(t-\tau_p)^2/(2\sigma^2)}.$
}
\label{cj_sho3lgaussprb}
\end{figure}

Now we consider a simple harmonic oscillator system with ground state and the first two excited states in a Gauss pulse laser field. For simplicity, the pump pulse width (the variance) and the Stokes pulse width are set to be the same ($\sigma$); the peak point of the Stokes pulse ($\tau_s$) and the peak point of the pump pulse ($\tau_p$) are set to be symmetric about the mid-point of the whole external field length. The pulse delay is $\Delta\tau=\tau_p-\tau_s$. Thus, the pulse shape functions $s \left(t\right)$ and $p\left(t\right)$ can be written as,

\begin{eqnarray}
s \left( t \right) &=& \frac{1}{\sigma \sqrt{2 \pi}}e^{-\frac{(t-\tau_s)^2}{2\sigma^2}},\\
p \left( t \right) &=& \frac{1}{\sigma \sqrt{2 \pi}}e^{-\frac{(t-\tau_p)^2}{2\sigma^2}}.
\end{eqnarray}

Then the Rabi requencies for the Stokes and pump pulses are,

\begin{eqnarray}
\Omega_s \left( t \right) &=&A_0 \frac{1}{\sigma \sqrt{2 \pi}}e^{-\frac{(t-\tau_s)^2}{2\sigma^2}},\\
\Omega_p \left( t \right) &=&A_0 \frac{1}{\sigma \sqrt{2 \pi}}e^{-\frac{(t-\tau_p)^2}{2\sigma^2}}.
\end{eqnarray}

We calculate the STIRAP process with different pulse variances and different time delays. In our calculation, the pulse delay $\Delta\tau$ varies from $0.7 \times 10^6$ to $1.6 \times 10^6$. For small pulse delays, $\Delta\tau=0.7 \sim 1.0 \times 10^6$, the pulse width varies from $1.0 \times 10^6$ to more than $3.0 \times 10^6$, while for large pulse delays, $\Delta\tau=1.1 \sim 1.6 \times 10^6$, the pulse width varies from $1.55 \times 10^6$ to more than $3.0 \times 10^6$. Fig.~(\ref{cj_sho3lgausssumprb} shows the target state population vs. pulse variance curves for different pulse delays. From the plot, we can see that the population transfer to the target state is related to the pulse variance $\sigma$ and pulse delay $\Delta\tau$. From the plot, it can be seen that for the same pulse delay, the population in the target state increases as the variance increases and then goes down slowly after it reaches a maximum value. The maximum value of the population in the target state for each time delay is larger than $0.88$ although the maximum points for different pulse delays are with different pulse widths (variance). While the pulse delay $\Delta\tau$ increase from $0.7 \times 10^6$ to $1.6 \times 10^6$, the pulse widths ( the pulse variance, $\sigma$) which leads to the maximum population transfer to the target state moves from $1.55 \times 10^6$ to $2.25 \times 10^6$.   

Fig.~(\ref{cj_sho3lgausssumprbpart}) is part of Fig.~(\ref{cj_sho3lgausssumprb}) in which only the neighbor part of the maximum populations is shown. From our calculating result, we obtained maximum target population $0.88889175$, which corresponds to pulse delay $\Delta\tau = 0.9 \times 10^6$ and pulse width $\sigma=1.4 \times 10^6$. Although this result is good enough to reflect the relationship between the population transfer and the pulse shap functions, calculation for more values of  pulse delay $\Delta\tau$ and pulse width $\sigma$ is needed to get more accurate values. 

Fig.~(\ref{cj_sho3lgausslaser}), Fig.~(\ref{cj_sho3lgaussomega}) and Fig.~(\ref{cj_sho3lgaussprb}) show the plots for the external laser field, the Rabi frequency, and population transfer. In these plots, the pulse delay is $\Delta\tau=0.9 \times 10^6$ and the pulse width is $\sigma=1.4 \times 10^6$, which are the maximum values from our calculation.

\section{Conclusion and Analysis}
%section V
For a simple harmonic oscillator, the energy differences $E_2 - E_1 = E_1-E_0$. When coherent STIRAP happens, the pump pulse frequency $\omega_p=\frac{E_1-E_0}{\hbar}$ and the Stokes pulse frequency $\omega_s=\frac{E_2-E_1}{\hbar}$ have same value $\omega_0$. Thus, the stimulated population transfer from energy state $\vert 1 \rangle$ to $\vert 0 \rangle$ happens when the pump pulse exists. This leads to the result that the population is not fully transferred to the second excited state.

When Gauss pulses are used as the external laser field, the population transfer in a STIRAP process of  a three level simple harmonic oscillator system can be obtained as much as $0.88889175$.

\section{Acknowledgement}
%section VI
This work was supported by the US Army Night Vision Laboratory.

\section{Appendix Chebyshev Polynomial}
%section V

The Chebyshev Polynomials are defined \cite{TRIVLIN90B} as.

\begin{equation}
T_n\left( \tau \right)=\cos n\theta,
\end{equation}

where

\begin{equation}
\tau=\cos \theta.
\end{equation}

The Chebyshev Polynomials are orthogonal over the interval $[-1,1]$,

\begin{equation}
\frac{2}{\pi}\int_{-1}^{1} \frac{T_n \left(\tau\right) T_m \left(\tau\right) }{\sqrt{1-{\tau}^2}} \mathrm{d} \tau = \delta_{n m}\left( 1 + \delta_{n 0} \right).
\end{equation}

The $N$ roots of the $Nth$ Chebyshev polynomial $T_N \left( x \right)$:

\begin{equation}
\tau_{k} = \cos \left( \frac{\pi \left( k + \frac{1}{2} \right) }{N} \right), \qquad n=0,1,\dots, N-1
\end{equation}

The orthogonal relations of the Chebyshev polynomials at the root points are

\begin{equation}
\frac{2}{N}\sum_{k=0}^{N-1}{T_n \left(\tau\right) T_m \left(\tau\right) } = \delta_{n m}\left( 1 + \delta_{n 0} \right).
\end{equation}

A function $f\left( \tau \right)$ can be expanded by the Chebyshev polynomial as

\begin{equation}
f \left(\tau \right)=\sum_{n=0}^{N-1} F_n {T_n \left( \tau \right)}.
\end{equation}

At the root points $\tau_k$,

\begin{equation}
f \left(\tau_k\right)=\sum_{n=0}^{N-1} F_n {T_n \left(\tau_k\right)}.
\end{equation}

The above equation multiplied by $ T_m \left( \tau_k \right)$ and take summation, 

\begin{equation}
\sum_{k=0}^{N-1} f \left(\tau_k\right)T_m \left( \tau_k \right)=\sum_{k=0}^{N-1} \sum_{n=0}^{N-1} F_n T_n \left(\tau_k\right) T_m \left(\tau_k\right).
\end{equation}

Thus $F_m$ can be obtained,

\begin{equation}
F_m = \frac{2 - \delta_{m 0} }{N}\sum_{k=0}^{N-1} f \left(\tau_k\right)T_m \left( \tau_k \right).
\end{equation}

So $f \left( \tau \right)$ can be expressed as,

\begin{equation}
f \left( \tau \right) = \sum_{k=0}^{N-1} f \left(\tau_k\right) \sum_{n=0}^{N-1} \frac{2  - \delta_{n 0} }{N} T_n \left( \tau_k \right)T_n \left( \tau \right).
\end{equation}

Take integration,

\begin{eqnarray}
& &\int_{-1}^{\tau^{\prime}} f \left( \tau \right) \mathrm{d} \tau \nonumber\\
&=& \sum_{k=0}^{N-1} f \left(\tau_k\right) \sum_{n=0}^{N-1} \frac{2  - \delta_{n 0} }{N} T_n \left( \tau_k \right)
\int_{-1}^{\tau^{\prime}} T_n \left( \tau \right) \mathrm{d} \tau.
\end{eqnarray}

Then any time integral can be obtained from

\begin{equation}
\int_{-1}^{\tau^{\prime}} f \left( \tau \right) \mathrm{d} \tau = \sum_{k=0}^{N-1} f \left(\tau_k\right) J_k \left( \tau^{\prime} \right),
\end{equation}

Where

\begin{equation}
J_k \left( \tau \right) = \sum_{n=0}^{N-1} \frac{2  - \delta_{n 0} }{N} T_n \left( \tau_k \right) S_n \left( \tau \right),
\end{equation}

and $S_n \left( \tau^{\prime} \right)$ is,
\begin{equation}
S_n \left( \tau^{\prime} \right) =\int_{-1}^{\tau^{\prime}} T_n \left( \tau \right) \mathrm{d} \tau.
\end{equation}

\end{document}